\documentclass[aps,pra,reprint,superscriptaddress,nofootinbib,longbibliography]{revtex4-2}

\usepackage{graphicx}
\usepackage{amsmath,amssymb}
\usepackage{booktabs}
\usepackage{algpseudocode}
\usepackage[colorlinks=true,allcolors=blue]{hyperref}
\usepackage{url}

\graphicspath{{figures/}}

\newcommand{\CVaR}{\mathrm{CVaR}}
\providecommand{\ket}[1]{|#1\rangle}
\newcommand{\ibmfez}{\texttt{ibm\_fez}}
\newcommand{\Ntwoq}{N_{2q}}
\newcommand{\Dtwoq}{D_{2q}}

\begin{document}

\title{Batched pattern search for QAOA parameter optimization on
cloud-accessed quantum hardware}

\author{Muhammad Faryad}
\email{muhammad.faryad@lums.edu.pk}
\affiliation{Department of Physics, Lahore University of Management Sciences (LUMS),
Lahore 54792, Pakistan}

\date{\today}

\begin{abstract}
On a cloud-accessed quantum processor the classical optimizer of a variational algorithm
communicates with the device through submitted jobs, and each job carries a queueing and
turnaround overhead that does not depend on how many circuits it contains. We study a
simple consequence of this for the quantum approximate optimization algorithm (QAOA). An
optimizer whose next set of trial parameters is known before any result returns can
evaluate the whole set in one job, whereas a sequential optimizer such as COBYLA spends
one job per function evaluation. We compare a batched pattern search with COBYLA on IBM's
\ibmfez\ processor for cardinality-constrained portfolio selection with six to twelve
assets, giving both optimizers the same number of jobs and the same number of shots. At
every size the batched search reaches, after its first job, a parameter quality that the
serial optimizer takes several jobs to match, and the two methods converge to comparable
final values; repeating the eight-asset comparison from four starting points gives the same
ordering each time. The instances are small enough to be solved exactly, which lets us
check that the device's output distribution is correlated with the true ranking of
portfolios and is clearly separated from a control circuit of the same depth with the cost
layer removed. A short scan over circuit depth at classically optimal parameters shows
that on this device the measured solution quality peaks at two or three QAOA layers. The
protocol is described in enough detail to be reproduced, and the notebooks and raw counts
are released.
\end{abstract}

\maketitle

\section{Introduction}
\label{sec:intro}

A variational quantum algorithm consists of a classical optimizer wrapped around a
parameterized quantum circuit~\cite{farhi2014qaoa,zhou2020qaoa,blekos2024qaoareview}, and
QAOA in particular has been run on superconducting processors at a range of
sizes~\cite{harrigan2021qaoa,weidenfeller2022scaling}. When
the circuit runs on a processor reached through a cloud queue, every exchange between the
optimizer and the device is a submitted job, and a job has costs that do not depend on
what it contains: it waits in a queue, it is loaded onto the control electronics, and its
results travel back. In the runs reported here the shortest round trip we observed on
IBM's \ibmfez\ was about $11$\,s and the median about $19$\,s, and the shortest did not
depend on what the job carried. An optimizer that submits one circuit evaluation per
job pays this overhead once per evaluation. Queueing and turnaround on shared quantum
hardware have been characterized before~\cite{ravi2021cloud,ravi2021adaptive,wu2024latency},
and their consequences for stochastic optimizers have been analyzed
theoretically~\cite{menickelly2023latency}. Our aim here is narrower and more practical: to
compare two classical optimizers for QAOA on real hardware when the number of submitted
jobs, rather than the number of circuit evaluations, is the quantity held fixed.

The comparison rests on a simple observation. A single job may carry many circuits ---
in Qiskit's primitive interface one primitive unified bloc (PUB) can hold an arbitrary
number of parameter vectors for the same circuit~\cite{javadiabhari2024qiskit} --- but an
optimizer can use that capacity only to the extent that it can name its next trial points
before it sees any result. Methods differ in how many points that is.
COBYLA~\cite{powell1994cobyla}, Nelder--Mead~\cite{nelder1965simplex} and Powell's
method~\cite{powell1964efficient} are sequential once past their initial simplex or
direction set, since each new point depends on the value just returned.
SPSA~\cite{spall1992spsa} names two points per iteration, and a parameter-shift
gradient~\cite{mitarai2018qcl,schuld2019gradients} names all of its evaluations at once.
Pattern search~\cite{torczon1997convergence,kolda2003directsearch,hough2001apps} names its
whole iteration in advance: the current point, a fixed stencil of trial points around it,
and, in the variant used here, one probe along the last accepted step. For a QAOA circuit
with $p$ layers that is $4p+2$ points, and all of them can be submitted as the rows of one
job.

We compare this batched pattern search with COBYLA on four instances of
cardinality-constrained portfolio selection with $6$, $8$, $10$ and $12$ assets, executed
on \ibmfez, with both optimizers given the same number of jobs and the same total number of
shots. COBYLA is used as the serial reference because it was the strongest of the common
sequential methods in the simulations we ran before using any hardware time
(Appendix~\ref{app:tuning}). The instances are small enough that the exact optimum and the
exact ranking of every feasible portfolio are known by enumeration, so the quality of any
parameter vector can be evaluated exactly and the device output can be compared with the
true ordering. Two controls accompany each run: a circuit of identical depth with the cost
layer switched off, which measures the noise floor, and a repeat of the first job at the
end of the run, which bounds drift. Two smaller follow-up experiments --- a scan over
circuit depth at classically optimal parameters, and a repeat of the $8$-asset comparison
from different starting points --- address questions raised by the main runs.

The results are what the counting argument leads one to expect, and their value, such
as it is, lies in having been measured on hardware with the controls in place. The batched
search reaches a good region of parameter space in one or two jobs where COBYLA needs
several; the two methods reach comparable final quality; and the device output at the
returned parameters is correlated with the exact ranking of portfolios and well separated
from the control. We also find that on this device the measured solution quality peaks at
two or three QAOA layers and declines beyond that, which is useful to know when choosing
the depth. None of this constitutes an advantage over classical solvers, which handle
these instances instantly; the sizes were chosen so that every claim could be checked
against an exact answer, not for difficulty.

The paper is written so that the experiment can be followed and repeated.
Section~\ref{sec:problem} states the optimization problem and its Ising form,
Sec.~\ref{sec:circuit} constructs the circuit, Sec.~\ref{sec:cost} describes what a job
costs and how the runs were budgeted, Sec.~\ref{sec:optimizer} states the optimizer,
Sec.~\ref{sec:protocol} fixes the protocol and its controls, Sec.~\ref{sec:results}
reports the hardware results, and Sec.~\ref{sec:discussion} discusses their limitations.
Appendix~\ref{app:tuning} records how the hyperparameters were chosen in simulation and
Appendix~\ref{app:repro} describes the released material and some implementation pitfalls.

\section{The problem and its Ising form}
\label{sec:problem}

We solve the discrete, cardinality-constrained form of Markowitz mean--variance portfolio
selection~\cite{markowitz1952portfolio}, a common test problem for quantum
optimization~\cite{egger2020finance,herman2023finance,hodson2019portfolio,buonaiuto2023portfolio}.
Let $x_i \in \{0,1\}$ indicate that asset $i$ is held, let $\mathbf r$ be the vector of
expected returns and $\Sigma$ the covariance matrix. With a hard budget of exactly $B$
assets, the cost function is
\begin{equation}
  C(\mathbf x) \;=\; \lambda\, \mathbf x^{\mathsf T}\Sigma\,\mathbf x
  \;-\; \mathbf r^{\mathsf T}\mathbf x
  \;+\; \mu\Big(\textstyle\sum_i x_i - B\Big)^{2},
  \label{eq:costfn}
\end{equation}
where the last term is the usual quadratic penalty that turns the constraint into a
QUBO~\cite{lucas2014ising,glover2019qubo}. Two coefficients in Eq.~\eqref{eq:costfn}
affect whether the problem is meaningful, and we describe how each was set.

\emph{The risk aversion $\lambda$.} If $\lambda$ is small the quadratic term is
irrelevant and the problem reduces to sorting assets by expected return. We fix $\lambda$
so that, over the feasible set, the range of the risk term equals the range of the return
term,
\begin{equation}
  \lambda \;=\;
  \frac{\max_{\rm feas}(\mathbf r^{\mathsf T}\mathbf x) - \min_{\rm feas}(\mathbf r^{\mathsf T}\mathbf x)}
       {\max_{\rm feas}(\mathbf x^{\mathsf T}\Sigma\mathbf x) - \min_{\rm feas}(\mathbf x^{\mathsf T}\Sigma\mathbf x)} ,
  \label{eq:lambda}
\end{equation}
so that the optimum is a trade-off between the two terms rather than a ranking by one of
them.

\emph{The penalty $\mu$.} If $\mu$ is too small the global minimum of
Eq.~\eqref{eq:costfn} is infeasible; if it is too large the penalty dominates every
Hamiltonian coefficient, the $ZZ$ couplings become nearly uniform, and the ansatz spends
its layers enforcing the Hamming weight rather than discriminating among feasible
portfolios. We scanned $\mu$ against the exactly optimized noiseless $p=2$ QAOA
(Appendix~\ref{app:tuning}) and took the value at which the noiseless approximation ratio
was largest, which is about $5\%$ of the spread of the penalty-free objective over the
feasible set. At all four sizes this is comfortably above the smallest penalty that makes
the ground state feasible.

Under $x_i = (1-z_i)/2$, Eq.~\eqref{eq:costfn} becomes an Ising Hamiltonian
\begin{equation}
  \tilde H \;=\; \sum_i h_i Z_i + \sum_{i<j} J_{ij} Z_i Z_j ,
  \qquad C \;=\; s\,\tilde H + c ,
  \label{eq:ising}
\end{equation}
where the identity term has been dropped and the coefficients rescaled so that
$\max(|h|,|J|) = 1$. The rescaling matters in practice because the pattern-search step
size and the shot-noise tolerance of Sec.~\ref{sec:optimizer} are expressed in these
units, and without it they would have to be retuned for every instance. In these units the
uniform superposition has $\langle\tilde H\rangle = 0$, so a run that ends at a positive
value has done worse than the trivial state. Each notebook checks that the diagonal of
Eq.~\eqref{eq:ising} reproduces Eq.~\eqref{eq:costfn} on all $2^n$ bitstrings before any
circuit is built.

The four instances are nested: instance $n$ consists of the first $n$ assets of a fixed
twelve-asset table of large-capitalization US equities, with $B = n/2$. Any change in the
results with $n$ is therefore a property of the same problem growing rather than of four
unrelated problems. Their parameters and exact solutions are listed in
Table~\ref{tab:instances}.

\begin{table}[t]
\caption{The four nested instances. $r$(uniform) is the expected approximation ratio of
the uniform distribution over the feasible set, which any method must exceed to have done
anything. All quantities are exact, from enumeration of all $2^n$ bitstrings.}
\label{tab:instances}
\begin{ruledtabular}
\begin{tabular}{cccccccc}
$n$ & $B$ & $\lambda$ & $\mu$ & $\binom{n}{B}$ & $C_{\min}$ & $C_{\max}^{\rm feas}$
 & $r$(uniform) \\
\hline
6 & 3 & 1.771 & 8 & 20 & $-136.8$ & 21.1 & 0.451 \\
8 & 4 & 1.653 & 9 & 70 & $-161.1$ & 23.3 & 0.456 \\
10 & 5 & 1.553 & 12 & 252 & $-219.4$ & 23.7 & 0.490 \\
12 & 6 & 1.360 & 13 & 924 & $-243.3$ & 26.0 & 0.505 \\
\end{tabular}
\end{ruledtabular}
\end{table}

We report three quantities. The \emph{feasible probability mass} is the fraction of shots
with Hamming weight $B$; post-selecting on it is a free form of error mitigation, since
every other bitstring is certainly an error. The \emph{expected approximation ratio
conditioned on feasibility} is
\begin{equation}
  r \;=\; \frac{C_{\max}^{\rm feas} - \langle C\rangle_{\rm feas}}
                {C_{\max}^{\rm feas} - C_{\min}} ,
  \label{eq:ratio}
\end{equation}
an average over the whole post-selected distribution, which, unlike a best-of-$N$ figure,
cannot be improved simply by taking more shots. The third is
$p(\text{optimum}\mid\text{feasible})$, the probability of sampling the exact optimum
among feasible outcomes, which we report as a multiple of the uniform value
$1/\binom{n}{B}$.

The optimizer does not minimize Eq.~\eqref{eq:ratio} directly. It minimizes the
conditional value at risk~\cite{barkoutsos2020cvar}, $\CVaR_\alpha$, the mean energy of the
lowest $\alpha$ fraction of the sampled probability mass, with $\alpha = 0.5$. On a noisy
device the mean energy is dominated by the depolarized background while the low-energy
tail still carries the signal; $\alpha = 0.5$ was the best of the three values tried in
simulation (Appendix~\ref{app:tuning}); a shorter tail carries less of the background but
more shot noise. Related tail-based objectives are discussed in Ref.~\cite{amaro2022filtering}.

\section{Circuit synthesis on a chain}
\label{sec:circuit}

The cost Hamiltonian of Eq.~\eqref{eq:ising} couples every pair of assets, so its
interaction graph is complete, while the processor's coupling map is a heavy-hexagonal
lattice. Handing the ansatz to a preset transpiler produces a routing solution whose depth
is set by a SWAP heuristic and varies with the random seed, which is inconvenient when the
circuit is meant to be the fixed, independent variable of an experiment.

We instead place the $n$ qubits on a single physical chain and realize all $\binom{n}{2}$
couplings with an odd--even transposition
network~\cite{kivlichan2018lineardepth,ogorman2019swapnetworks,weidenfeller2022scaling}:
at each of $n$ steps, disjoint neighbouring pairs interact and then swap, so that after
$n$ steps every pair has met exactly once. Writing $\mathrm{RZZ}(\theta)\cdot\mathrm{SWAP}$
explicitly as
\begin{equation}
  \texttt{cx}(a,b)\;\texttt{rz}(\theta,b)\;\texttt{cx}(b,a)\;\texttt{cx}(a,b),
  \label{eq:rzzswap}
\end{equation}
which is exact up to a global phase, costs three entangling gates per pair. The resulting
gate count is deterministic,
\begin{equation}
  \Ntwoq = 3p\binom{n}{2}, \qquad \Dtwoq = 3pn ,
  \label{eq:gatecount}
\end{equation}
independent of the transpiler seed, and the final permutation of qubits is tracked so that
classical bit $i$ always carries logical qubit $i$.

\begin{figure}[t]
\includegraphics[width=\columnwidth]{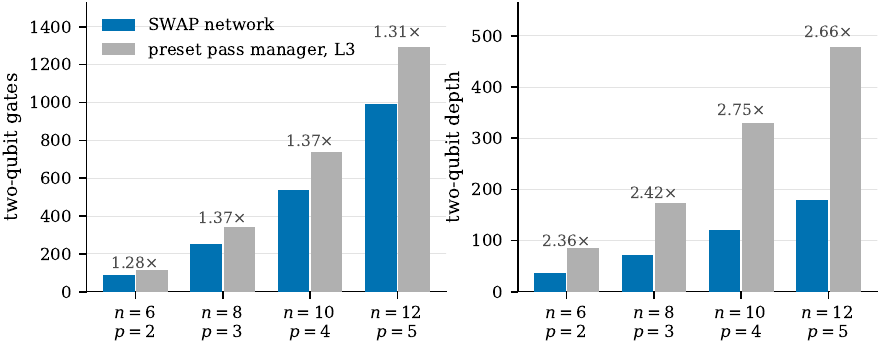}
\caption{Two-qubit gate count and two-qubit depth of the swap-network synthesis compared
with Qiskit's preset pass manager at optimization level~3, for the four circuits actually
submitted. Labels give the ratio.}
\label{fig:circuit}
\end{figure}

Emitting Eq.~\eqref{eq:rzzswap} by hand rather than leaving it to the compiler is worth
the small effort: Qiskit's two-qubit block consolidation does not fuse gates that carry
unbound parameters, so a transpiled $\mathrm{RZZ}\cdot\mathrm{SWAP}$ costs five entangling
gates instead of three. Table~\ref{tab:circuit} and Fig.~\ref{fig:circuit} compare the
hand-built network with the preset pass manager at optimization level~3 for the four
circuits actually submitted; the network uses somewhat fewer two-qubit gates and is
between two and three times shallower. The chain itself is chosen at run time by a beam
search over simple paths of the coupling map that minimizes $\sum -\log(1-\varepsilon)$
over the two-qubit and readout errors along the path, so that the circuit sits on the best
line the device offers on the day, and the notebook checks that the pass manager left the
layout in place. Each circuit is verified against the exact QAOA state vector before
submission.

\begin{table}[t]
\caption{Circuit resources on \ibmfez. The swap-network columns follow
Eq.~\eqref{eq:gatecount}; the preset columns depend on the transpiler seed.}
\label{tab:circuit}
\begin{ruledtabular}
\begin{tabular}{cccccccc}
& & \multicolumn{2}{c}{swap network} & \multicolumn{2}{c}{preset, level 3}
 & \multicolumn{2}{c}{ratio} \\
$n$ & $p$ & $N_{2q}$ & $D_{2q}$ & $N_{2q}$ & $D_{2q}$ & gates & depth \\
\hline
6 & 2 & 90 & 36 & 115 & 85 & 1.28 & 2.36 \\
8 & 3 & 252 & 72 & 344 & 174 & 1.37 & 2.42 \\
10 & 4 & 540 & 120 & 738 & 330 & 1.37 & 2.75 \\
12 & 5 & 990 & 180 & 1293 & 478 & 1.31 & 2.66 \\
\end{tabular}
\end{ruledtabular}
\end{table}

\section{What a job costs}
\label{sec:cost}

IBM defines the metered usage of a job as the time for which a QPU is locked to execute
it~\cite{ibmdocs_runtime}. The documentation gives an estimate of the form
\emph{(per-sub-job overhead) $+$ (rep\_delay $+$ circuit length) $\times$ (number of
executions)}, with an overhead of roughly two seconds per sub-job. Every job of this study
submitted a single PUB in job mode --- no \texttt{Session} and no \texttt{Batch} --- which
is the access mode available on the free Open Plan~\cite{ibmdocs_plans}.

Instantiating that estimate for \ibmfez, with its $250\,\mu$s default repetition delay and
the gate durations from its calibration data, and rounding the overhead up to three
seconds for budgeting, gives
\begin{equation}
  T_{\rm QPU} \simeq 3\,\text{s} + \big(352.6 + 0.267\,\Dtwoq\big)\,\mu\text{s}
                \times N_{\rm shots},
  \label{eq:cost}
\end{equation}
with $\Dtwoq$ the two-qubit depth. The per-shot coefficient is dominated by the repetition
delay and readout; the depth-dependent part is small for the circuits used here. Equation
\eqref{eq:cost} is what the ledger described in Appendix~\ref{app:repro} used to budget
each run in advance.

We had also intended to record the runtime's own reported usage for every job, and this
did not work: the key path used to read \texttt{job.metrics()} does not exist in the
client version we ran, and the surrounding \texttt{try/except} fell back to the predicted
value without warning. We discovered this only when auditing the released data, in which
the recorded and predicted usage are consequently identical. Every figure for metered QPU
time in this paper is therefore Eq.~\eqref{eq:cost} evaluated on the job's contents, not
a measurement of the device. We expect it to be close, since it follows IBM's own
estimator, but no conclusion below depends on it, and we say so wherever it is quoted.

\begin{figure*}[t]
\includegraphics[width=\textwidth]{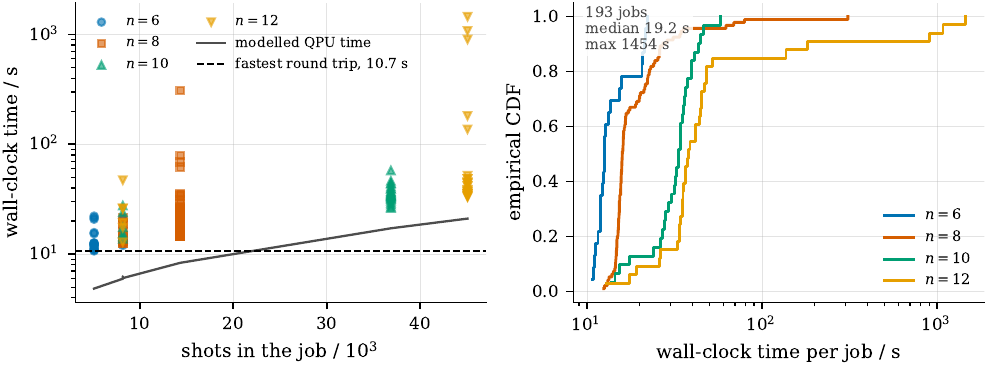}
\caption{The cost of a job on \ibmfez, from the $193$ jobs of this study. \textbf{Left:}
measured wall-clock time from submission to result against the number of shots the job
carried. The dashed line marks the shortest round trip observed. \textbf{Right:} the
distribution of the same times, by instance size. Both panels are measured; the estimated
metered time of Eq.~\eqref{eq:cost} is shown for scale only.}
\label{fig:cost}
\end{figure*}

What was measured for every job is the wall-clock time from submission to result. Across
the $193$ jobs of this study the median round trip was about $19$\,s, the shortest about
$11$\,s, and the distribution has a long tail with one job taking about $24$ minutes
(Fig.~\ref{fig:cost}). There is, in other words, a floor of order ten seconds on a round
trip that no reduction in the contents of a job removes, and over all $193$ jobs the
wall-clock time exceeded the estimated metered time by a factor of about four (by more
for the four main runs alone; Table~\ref{tab:ledger}). Table~\ref{tab:ledger} lists the
resources used by the four main runs; the depth scan and the replicate study added $25$
and $66$ jobs respectively.

\begin{table*}[t]
\caption{Resource ledger for the four main runs. ``QPU'' is Eq.~\eqref{eq:cost} evaluated
on each job's contents, an estimate rather than a measurement for the reason given in the
text; ``wall'' is the measured sum of submission-to-result times. The two follow-up studies
added a further $91$ jobs. Every job identifier is in the released data.}
\label{tab:ledger}
\begin{ruledtabular}
\begin{tabular}{cccccccccc}
$n$ & $p$ & shots/cand. & shots/job & jobs & total shots & QPU\,/\,s
 & wall\,/\,s & ratio & median wall\,/\,s \\
\hline
6 & 2 & 512 & 5{,}120 & 20 & 111{,}600 & 100 & 292 & 2.9 & 13 \\
8 & 3 & 1024 & 14{,}336 & 24 & 325{,}612 & 193 & 942 & 4.9 & 23 \\
10 & 4 & 2048 & 36{,}864 & 28 & 946{,}160 & 448 & 968 & 2.2 & 34 \\
12 & 5 & 2048 & 45{,}056 & 30 & 1{,}241{,}066 & 587 & 4688 & 8.0 & 38 \\
\hline
total & & & & 102 & 2{,}624{,}438 & 1329 & 6890 & 5.2 & \\
\end{tabular}
\end{ruledtabular}
\end{table*}

The consequence for a variational loop is the same under either accounting. An optimizer
that submits one job per function evaluation pays the per-job overhead once per
evaluation, before any shot is taken; a batched iteration that puts $M$ parameter vectors
into one PUB pays it once for all $M$. At the settings used below $M$ is between $10$ and
$22$.

Two qualifications belong here. First, wall-clock time is not a property of the device or
of the algorithm: it depends on other users' load and on fair-share priority, and our own
per-run ratios of wall-clock to estimated metered time vary by a factor of about four
between runs (Table~\ref{tab:ledger}). It should be read as a characterization of the
access mode on the days we used it, not as a constant. Second, that access mode is job mode on the free plan. IBM's \texttt{Batch} and
\texttt{Session} modes exist precisely to amortize queueing across a sequence of jobs, and
a user with access to them would see a smaller wall-clock penalty per job, though the
per-job metered overhead of Eq.~\eqref{eq:cost} is charged in all three
modes~\cite{ibmdocs_runtime}. The argument of this paper is therefore strongest for the
open-plan user and weaker, but not absent, for one with reserved capacity.

\section{Batched pattern search}
\label{sec:optimizer}

Pattern search, also called compass search or direct search, replaces derivatives with a
fixed stencil of trial points around an incumbent, accepting a move only when a trial point
improves on the incumbent and contracting the stencil when none
does~\cite{torczon1997convergence,kolda2003directsearch}. The property that matters here
is structural rather than statistical: the entire iteration is determined by the incumbent
and the current radius, so every point can be named before any value comes back. The
classical direct-search literature exploited this parallelism long
ago~\cite{hough2001apps}; we simply apply it to a job queue.

Algorithm~1 states the variant used. For $2p$ parameters an iteration consists of the
incumbent (row $0$), the $2\times 2p$ axial probes at radius $\rho$, and, when the previous
iteration moved, one momentum probe continuing along the accepted step: at most $4p+2$
points, all submitted as the rows of a single PUB in a single job.

\begin{figure}[t]
\vspace{2pt}\hrule\vspace{3pt}
\noindent\textbf{Algorithm 1.}~Batched pattern search. One iteration is one job.
\vspace{2pt}\hrule\vspace{4pt}
\begin{algorithmic}[1]
\Require incumbent $\mathbf x_0$, radius $\rho$, shot budget $S_{\rm job}$ per job
\State $\mathbf x \gets \mathbf x_0$, \; $\mathbf x_{\rm prev} \gets \mathbf x_0$
\For{$j = 1 \dots J$} \Comment{$J$ = the job budget}
  \State $\mathcal{C} \gets \{\mathbf x\} \cup
         \{\mathbf x \pm \rho\,\mathbf e_i\}_{i=1}^{2p}$
  \If{$\mathbf x \neq \mathbf x_{\rm prev}$}
     \State $\mathcal{C} \gets \mathcal{C} \cup
            \{2\mathbf x - \mathbf x_{\rm prev}\}$ \Comment{momentum probe}
  \EndIf
  \State submit \textbf{one job}: one PUB, $|\mathcal{C}|$ parameter rows,
         $S_{\rm job}/|\mathcal{C}|$ shots each
  \State $v_c \gets \CVaR_\alpha$ of the distribution measured for $c\in\mathcal{C}$
  \State $\sigma \gets$ standard deviation of $\tilde H$ under row $0$
  \State $\epsilon \gets \tfrac12\,\sigma/\sqrt{S_{\rm job}/|\mathcal{C}|}$
         \Comment{shot-noise tolerance}
  \State $c^\star \gets \arg\min_{c} v_c$
  \If{$c^\star \neq \mathbf x$ \textbf{ and } $v_{c^\star} < v_{\mathbf x} - \epsilon$}
     \State $\mathbf x_{\rm prev} \gets \mathbf x$, \;
            $\mathbf x \gets c^\star$, \;
            $\rho \gets \min(1.3\rho, \pi/2)$
  \Else
     \State $\rho \gets \max(0.6\rho, 0.02)$
  \EndIf
\EndFor
\end{algorithmic}
\vspace{2pt}\hrule\vspace{2pt}
\end{figure}

Two details deserve comment. The acceptance test requires an improvement larger than
$\epsilon$, half the standard error of the measured energy of the incumbent row. Without
it, at the shot counts used here, the incumbent moves on nearly every iteration and the
search follows shot noise rather than the landscape; the same measurement that supplies
the incumbent value supplies $\sigma$, so the tolerance costs nothing extra. The momentum
probe uses the fact that the batch has a spare row whenever the previous iteration moved,
and extrapolates along the accepted direction; it is a cheap substitute for a line search.

Because the number of candidates and the shots per job are both fixed in advance, the
metered cost of a run of $J$ iterations can be estimated from Eq.~\eqref{eq:cost} before
the run starts. A serial optimizer does not have this property: COBYLA's evaluation count is
data-dependent and has to be truncated at some chosen point.

The initial point is the linear ramp $\beta_k = \Delta t\,(1 - (k-\tfrac12)/p)$,
$\gamma_k = -\Delta t\,k/p$ with $\Delta t = 0.5$, an adiabatically motivated schedule of
the kind introduced in Ref.~\cite{zhou2020qaoa} and shown to avoid the poor local minima
that random initialization often finds~\cite{sack2021annealinginit}. Warm-starting from a
classical relaxation~\cite{egger2021warmstart} and parameter
transfer~\cite{akshay2021concentration,galda2021transfer,sureshbabu2024parameter} are
alternatives we did not use. The sign of $\gamma$ follows the convention of
Eq.~\eqref{eq:ising}.

\section{Protocol and controls}
\label{sec:protocol}

\begin{figure}[t]
\includegraphics[width=\columnwidth]{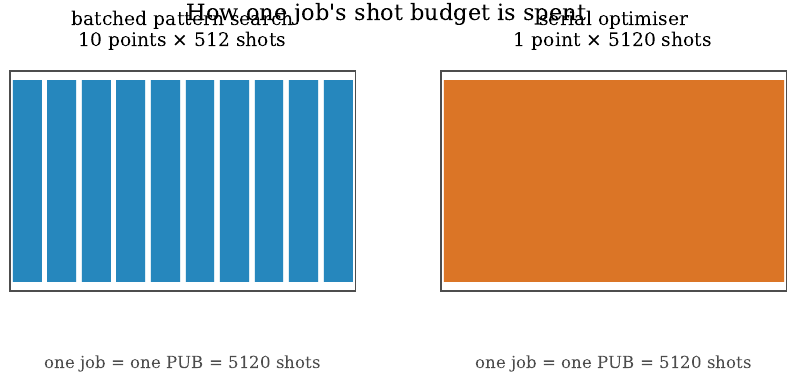}
\caption{How the two arms spend the shot budget of one job, drawn for the $6$-asset
settings. The batched arm divides the job's shots among $4p+2$ parameter vectors; the
serial arm spends all of them on one.}
\label{fig:concept}
\end{figure}

Each instance is run as a self-contained experiment with two arms under an identical
budget (Fig.~\ref{fig:concept}).

\emph{Matched budget.} Both arms receive the same number of jobs $J$ and the same shots
per job $S_{\rm job}$. The batched arm spends a job on $M = 4p+2$ parameter vectors at
$S_{\rm job}/M$ shots each; the serial arm spends a job on one parameter vector at the
whole $S_{\rm job}$. The two arms therefore consume the same quantum resources by both
measures, and the difference under test is how a job's shots are allocated.

This constraint nominally gives the serial arm $M$ times as many shots per evaluation,
or a $\sqrt{M}$-fold advantage in standard error. In practice the simulations of Appendix~\ref{app:tuning} found the final
quality insensitive to shots per candidate over the range used here, so a serial
evaluation at tens of thousands of shots is well into the regime where extra shots buy
little. We keep the constraint because it removes shot count as an alternative explanation
of the result, not because it is a large handicap for the batched arm.

One asymmetry runs the other way and should also be stated. The batched arm's $M$ rows
execute inside a single job under one calibration, so comparisons among them are free of
inter-job drift; COBYLA's successive evaluations sit in different jobs, minutes apart, and
carry that drift on top of shot noise. This favours the batched arm and is part of the
difference under test, but it means that difference is not purely one of allocation.

\emph{The serial optimizer.} COBYLA~\cite{powell1994cobyla} with $\rho_{\rm beg}$ equal
to the initial pattern-search radius. It was chosen because it was the strongest of
COBYLA, SPSA, Nelder--Mead and Powell in the simulations of Appendix~\ref{app:tuning} at
every size over the range of job budgets used here; the reference should be the best
available serial method rather than a weak one. (SPSA is stronger at a budget of two jobs
and Powell overtakes beyond about forty; neither is the regime studied.) This ordering is
consistent with published comparisons of optimizers under
noise~\cite{lavrijsen2020optimizers,bonetmonroig2023comparison}.

\emph{Error suppression.} Dynamical decoupling~\cite{viola1999decoupling,ezzell2023dd} on
idle qubits and Pauli twirling of gates and measurements~\cite{wallman2016twirling} were
enabled for every job; neither consumes extra shots. No zero-noise extrapolation or
probabilistic error cancellation~\cite{temme2017mitigation} was used, so apart from these
two settings the measured distributions are raw device output, and the only mitigation
applied anywhere in the paper is post-selection on Hamming weight.

\emph{The $\gamma=0$ control.} After the two optimization arms, one job is submitted at
the batched arm's best $\boldsymbol\beta$ with $\boldsymbol\gamma$ set to zero. The
circuit has the same gate count and depth, but its ideal output is exactly
$\ket{+}^{\otimes n}$. Any structure that survives in its output is the device's noise
floor rather than something the algorithm found, and every result below should be read
against it.

\emph{The drift control.} The two arms run one after the other, so a device that drifts
over the half hour of a run would favour whichever arm ran under better conditions. The
last job of every run repeats the first batched iteration exactly. The shift in measured
$\CVaR$ between the two was of the order of the shot noise and of either sign at the four
sizes, so no drift correction is applied.

\emph{Two views of quality.} Because $n \le 12$, the noiseless output distribution of any
parameter vector can be computed exactly. Alongside what the device measured we therefore
report the exact noiseless approximation ratio of the best point each arm holds after each
job. For the batched arm ``holds'' means the incumbent, the point that has survived the
acceptance test of Algorithm~1, and the curve is the running maximum of its noiseless
ratio. For the serial arm it is the running maximum over all the points COBYLA has
evaluated, an upper bound on what COBYLA would return. This metric is generous to the serial arm,
since the batched arm is credited only with points its acceptance rule kept while COBYLA
is credited with the best of everything it touched. It is also an oracle: neither
optimizer can compute it, and it is unavailable at any size that could not be simulated.
It is a diagnostic of the search rather than a quantity either method returns, and the
device read-outs of Sec.~\ref{sec:results} are its non-oracle counterpart.

\emph{Hyperparameters.} The objective ($\alpha$), the initial radius and the penalty
$\mu$ were chosen on a noise model of \ibmfez\ before any hardware time was spent, and
the same simulations guided the choice of shots and jobs (Appendix~\ref{app:tuning}). The
depth $p$ at $n = 10$ and $12$ was set by hand above the range simulated; the consequences
of this are examined in Sec.~\ref{sec:depth}.

\section{Results}
\label{sec:results}

\subsection{Convergence per job}

\begin{figure*}[t]
\includegraphics[width=\textwidth]{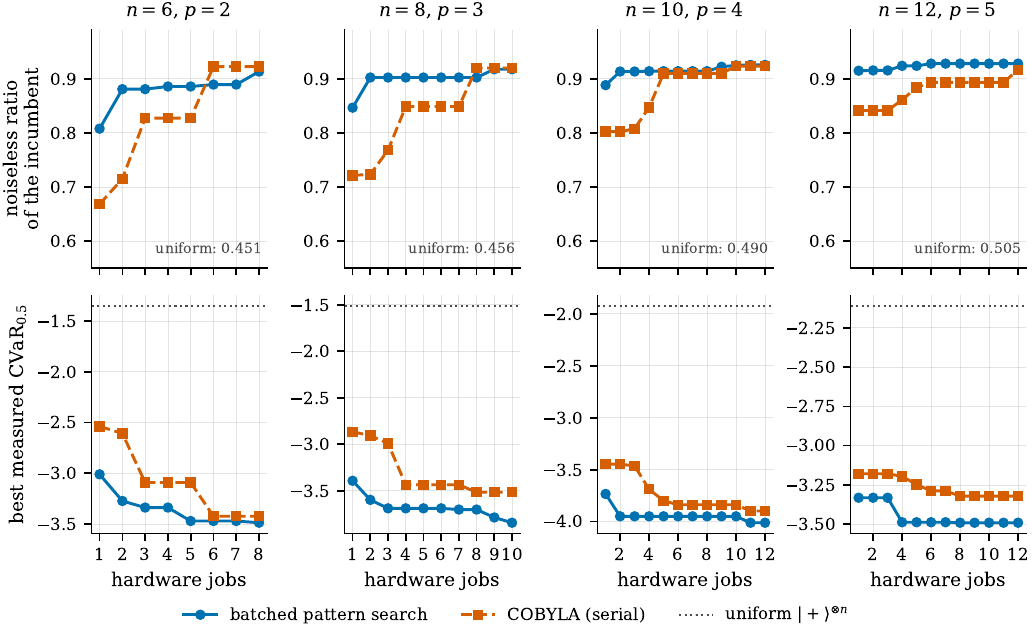}
\caption{Optimization on \ibmfez\ under a matched budget of jobs and shots.
\textbf{Upper:} running maximum of the exact noiseless approximation ratio of the
incumbent parameters, recomputed classically after every job; this is the quality of the
point the optimizer has found, independent of how well the device can realize it. \textbf{Lower:} the best
$\CVaR_{0.5}$ measured on the device so far. Both arms receive the same number of jobs and
the same shots per job; the batched arm spends a job on $4p+2$ parameter vectors, the
serial arm on one.}
\label{fig:convergence}
\end{figure*}

Figure~\ref{fig:convergence} shows the two arms at the four sizes and
Table~\ref{tab:convergence} summarizes it. The upper row is the running maximum of the
exact noiseless approximation ratio of each arm's incumbent, as a function of the number of
jobs spent; the lower row is what the device reported.

\begin{table*}[t]
\caption{Job efficiency on hardware. ``COBYLA needs'' is the number of jobs after which
the serial arm first reaches the noiseless quality the batched arm had after its first
job. ``Jobs to 95\%'' is the number of jobs each arm needs to reach $95\%$ of its own final
noiseless ratio. ``Best so far'' is the running maximum at the end of the budget, and
``returned'' the noiseless ratio of the parameters each arm actually returned for the
read-out of Table~\ref{tab:final}. At $n=12$ COBYLA matches the batched arm's one-job
quality only on the last job of its budget, and marginally.}
\label{tab:convergence}
\begin{ruledtabular}
\begin{tabular}{ccccccccccc}
& & & \multicolumn{2}{c}{after one batched job} & \multicolumn{2}{c}{jobs to 95\%}
 & \multicolumn{2}{c}{best so far} & \multicolumn{2}{c}{returned} \\
$n$ & $p$ & jobs/arm & $r_{\rm batched}$ & COBYLA needs & batched & COBYLA
 & batched & COBYLA & batched & COBYLA \\
\hline
6 & 2 & 8 & 0.808 & 3 & 2 & 6 & 0.913 & 0.922 & 0.913 & 0.922 \\
8 & 3 & 10 & 0.847 & 4 & 2 & 8 & 0.918 & 0.920 & 0.918 & 0.920 \\
10 & 4 & 12 & 0.888 & 5 & 1 & 5 & 0.926 & 0.923 & 0.926 & 0.883 \\
12 & 5 & 12 & 0.915 & 12 & 1 & 5 & 0.928 & 0.916 & 0.923 & 0.891 \\
\end{tabular}
\end{ruledtabular}
\end{table*}

The pattern is the same at all four sizes. After its first job the batched search already
holds parameters of a quality that COBYLA first reaches after several jobs, and measured
by the number of jobs needed to reach $95\%$ of its own final value, the batched search
takes one or two where COBYLA takes several more (Table~\ref{tab:convergence}). The number
of jobs COBYLA needs to catch up grows along the sequence, but we are cautious about
reading that as a size effect: the batch width $M = 4p+2$ and COBYLA's initial-model cost $2p+1$ both grow with
$p$, and $p$ was raised together with $n$, so instance size and batch width are confounded
in these runs. The statement the data support is the weaker one, that the ordering holds
at every size tested and, at $n = 8$, at every one of four starting points
(Sec.~\ref{sec:replicates}).

The lower row of Fig.~\ref{fig:convergence} is consistent with the following mechanism.
The serial traces are jagged, because COBYLA spends its early jobs building a linear model from single points
and several of those points are worse than where it started. The batched traces are
nearly monotone, because each job contains its own incumbent together with a stencil
around it, and the acceptance test rejects moves that are not larger than shot noise. The
batched search is not finding better optima --- measured by the best point either arm
reached, the two end within about $0.01$ of each other at every size --- it is finding
comparable ones in fewer round trips.

\subsection{What the device delivered}

\begin{figure*}[t]
\includegraphics[width=\textwidth]{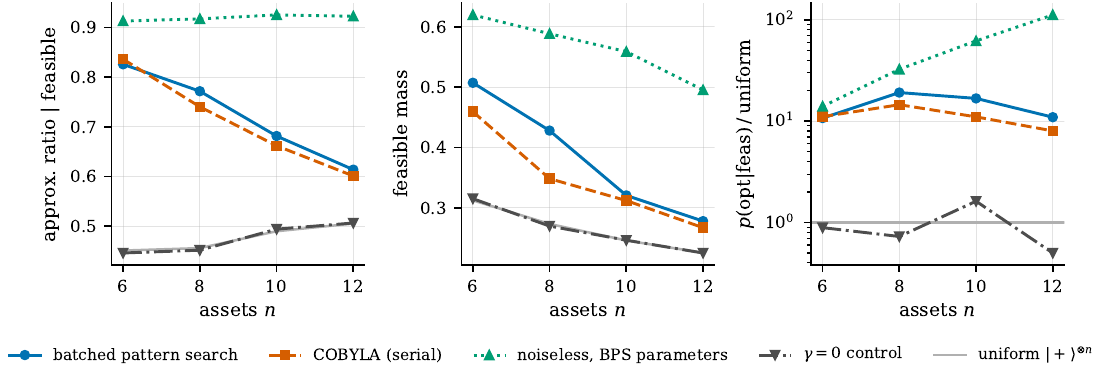}
\caption{Final read-out quality against instance size, at $8192$ shots. The $\gamma=0$
control runs a circuit of identical gate count and depth whose ideal output is
$\ket{+}^{\otimes n}$; it tracks the uniform baseline at every size, which is what allows
the separation between the measured curves and that baseline to be read as signal. The
right panel is on a logarithmic scale; the downward arrow on the $\gamma=0$ control at
$n=12$ marks an upper bound, that circuit having never sampled the optimum in $8192$
shots.}
\label{fig:final}
\end{figure*}

Figure~\ref{fig:final} and Table~\ref{tab:final} report the read-out at each arm's
returned parameters, at $8192$ shots, against the $\gamma=0$ control and the uniform
baseline.

\begin{table*}[t]
\caption{Final read-out on \ibmfez\ at each arm's returned parameters. $\rho_s$ is the
Spearman rank correlation between the measured probability of a feasible portfolio and its
exact cost; $\rho_s^{\gamma=0}$ is the same quantity for the noise-floor control.}
\label{tab:final}
\begin{ruledtabular}
\begin{tabular}{ccccccccccc}
& \multicolumn{4}{c}{approximation ratio $\mid$ feasible}
 & \multicolumn{2}{c}{feasible mass} & & & \multicolumn{2}{c}{rank corr.} \\
$n$ & batched & COBYLA & $\gamma{=}0$ & uniform & batched & uniform
 & $\frac{p({\rm opt}\mid{\rm feas})}{\rm uniform}$ & rank of opt.
 & $\rho_s$ & $\rho_s^{\gamma=0}$ \\
\hline
6 & 0.826 & 0.836 & 0.446 & 0.451 & 0.507 & 0.312 & 10.7 & 1/20 & $-0.97$ & $+0.07$ \\
8 & 0.772 & 0.740 & 0.451 & 0.456 & 0.428 & 0.273 & 19.2 & 1/70 & $-0.88$ & $+0.07$ \\
10 & 0.681 & 0.662 & 0.494 & 0.490 & 0.321 & 0.246 & 16.8 & 1/252 & $-0.74$ & $-0.08$ \\
12 & 0.614 & 0.601 & 0.506 & 0.505 & 0.278 & 0.226 & 11.0 & 2/924 & $-0.45$ & $-0.00$ \\
\end{tabular}
\end{ruledtabular}
\end{table*}

Three things can be read from the table. First, the device output is well separated from
its own noise floor at every size: the $\gamma=0$ control tracks the uniform baseline
closely in every case, while the optimized circuits sit well above it, by a margin that
narrows as $n$ grows. Second, the batched arm's read-out is better than the serial arm's
at $n = 8$, $10$ and $12$ and marginally worse at $n = 6$. At $n = 10$ and $12$ this
follows the noiseless quality of the returned parameters (last two columns of
Table~\ref{tab:convergence}); at $n = 6$ and $8$ the two parameter sets are noiselessly
equivalent, so the measured difference there reflects device noise rather than a property
of the optimizer, and the replicate study of Sec.~\ref{sec:replicates} is the appropriate
place to look for a systematic difference. Third, the probability of sampling the exact
optimum among feasible outcomes is enhanced over uniform by more than an order of
magnitude at every size.

The narrowing of the separation from the baseline with $n$ has two causes, and
Sec.~\ref{sec:depth} separates them: part of it is a consequence of having increased the
circuit depth along with the instance size, and part of it is the instance size itself.

\subsection{Rank correlation with the exact cost}

\begin{figure*}[t]
\includegraphics[width=\textwidth]{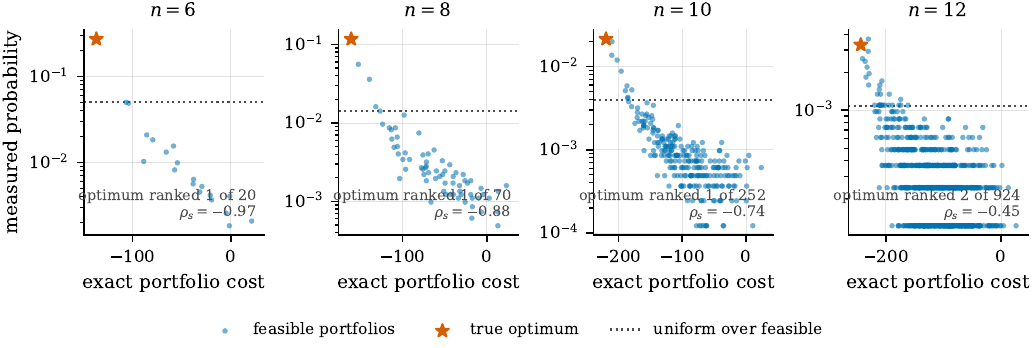}
\caption{Measured probability of every feasible portfolio against its exact cost, from
the batched arm's read-out. The star is the true optimum; $\rho_s$ is the Spearman rank
correlation. The corresponding correlation for the $\gamma=0$ control is consistent with
zero at every size (Table~\ref{tab:final}).}
\label{fig:ranking}
\end{figure*}

Figure~\ref{fig:ranking} plots, for every feasible portfolio, the probability the
device assigned it against its exact cost.

The relationship is monotone, strong at the smaller sizes and weaker at $n = 12$
(Table~\ref{tab:final}). The true optimum is the most probable feasible outcome at $n = 6$, $8$ and $10$, and the
second most probable of $924$ candidates at $n = 12$, where the shot noise on a few tens
of counts no longer resolves the leading few. The same statistic on the $\gamma=0$ control
is consistent with zero at every size, which is the check that makes the correlation
meaningful: a circuit of the same depth on the same qubits, differing only in that its
cost-layer angles are zero, carries no information about the cost function. The control thus excludes any artefact
of the readout chain or the gate set that is independent of the cost-layer angles.

\subsection{The device as a depolarizing channel}
\label{sec:noise}

The measured distributions are reasonably well described by a one-parameter model.
Writing $P_{\rm ideal}$ for the exact QAOA distribution at the measured parameters, we fit
\begin{equation}
  P_{\rm dev} \;=\; \mathcal{R}\Big[(1-\eta)P_{\rm ideal} + \eta\,\mathcal{U}\Big] ,
  \label{eq:twin}
\end{equation}
where $\mathcal{U}$ is the uniform distribution and $\mathcal{R}$ applies independent
readout bit flips at the calibrated rate, choosing $\eta$ so that the model reproduces the
measured mean energy: one fitted number per instance. The fitted $\eta$ rises from about
$0.25$ at $n = 6$ to about $0.78$ at $n = 12$ as the two-qubit gate count grows
(Table~\ref{tab:circuit}).

\begin{figure*}[t]
\includegraphics[width=\textwidth]{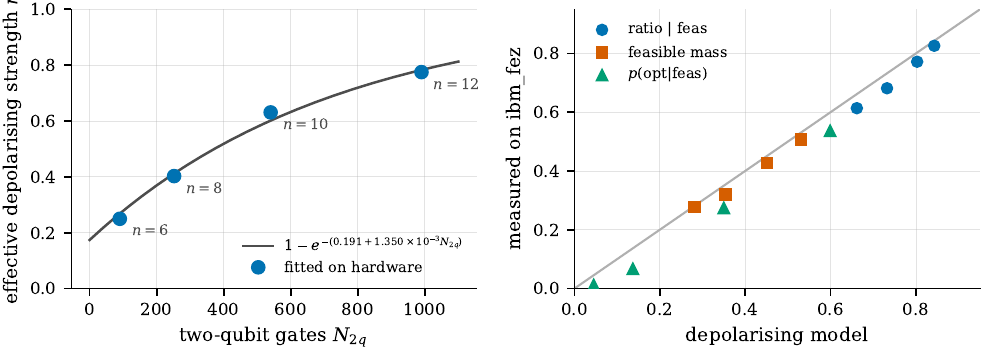}
\caption{\textbf{Left:} effective depolarizing strength $\eta$ fitted to the measured
mean energy of each instance, against the two-qubit gate count of its circuit; the line is
Eq.~\eqref{eq:etalaw}, a two-parameter fit to the four points. \textbf{Right:} the model of
Eq.~\eqref{eq:twin}, with $\eta$ fitted only to the mean energy, against the measured
values of three quantities it was not fitted to.}
\label{fig:noise}
\end{figure*}

The four values are described by
\begin{equation}
  \eta(\Ntwoq) \;=\; 1 - \exp\!\big[-(0.19 + 1.35\times 10^{-3}\,\Ntwoq)\big]
  \label{eq:etalaw}
\end{equation}
to within $0.03$ (Fig.~\ref{fig:noise}, left). The slope corresponds to an effective
error of $1.35\times10^{-3}$ per two-qubit gate, roughly half the median CZ error of
$0.26\%$ reported by the device calibration on the day. The two need not agree: a
depolarizing channel acting on the output distribution is a weaker statement than one
acting on the state, and Pauli twirling converts coherent errors into stochastic ones,
which are partly benign for a sampled distribution. The intercept is the depth-independent
contribution from state preparation, single-qubit gates and measurement.

The right panel of Fig.~\ref{fig:noise} compares the model with quantities it was not
fitted to. It reproduces the conditional approximation ratio and the feasible mass to
within about $0.06$ at every size, and under-predicts the loss in
$p(\text{optimum}\mid\text{feasible})$, as one would expect: a global depolarizing channel
damages the whole distribution uniformly, whereas correlated and coherent errors damage
the sharp peak more. Equation~\eqref{eq:etalaw} may therefore serve as a rough planning
estimate for this device, converting a proposed $(n,p)$ into an expected output quality,
and is somewhat optimistic as a physical description. With four points and two parameters
it should not be read as more than that.

\subsection{Depth scan at classically optimal parameters}
\label{sec:depth}

The circuit depth was increased along with the instance size in the runs above, which
leaves the two effects entangled. They can be separated without an optimizer. For
$n \le 12$ the noiseless QAOA landscape can be optimized exactly and classically, so for
each $(n,p)$ we computed the best parameters any optimizer could return at that depth and
then measured what the device made of them: one job of $8192$ shots per $(n,p)$, for
four sizes and five depths, plus a second parameter set at $n = 12$ described below. This
removes the optimizer from the picture and leaves only the depth--noise trade-off.

\begin{figure*}[t]
\includegraphics[width=\textwidth]{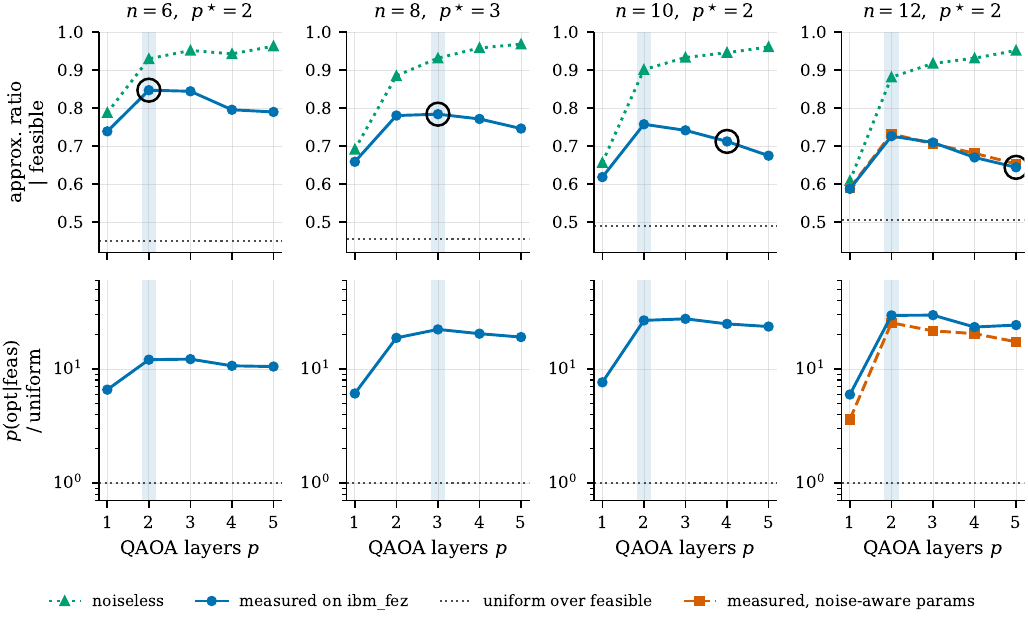}
\caption{Approximation ratio against circuit depth at classically optimal parameters. The
noiseless curve (green) rises with $p$; the measured curve (blue) rises, peaks at
$p^\star$ (shaded), and falls, because the $3p\binom{n}{2}$ two-qubit gates eventually
cost more fidelity than an extra layer gains. Black rings mark the depth actually used for
that size in the main runs. At $n=12$ the orange curve repeats the measurement at
parameters optimized against the noise model of Eq.~\eqref{eq:twin} instead of the
noiseless landscape. \textbf{Lower:} the probability of sampling the exact optimum,
conditioned on feasibility, as a multiple of the uniform value $1/\binom{n}{B}$.}
\label{fig:depth}
\end{figure*}

\begin{table}[t]
\caption{Depth scan on \ibmfez, $8192$ shots per point, at classically optimal parameters.
First row of each triple: measured approximation ratio. Second, in italic: the exact
noiseless value at the same parameters. Third: the measured
$p(\text{optimum}\mid\text{feasible})$ as a multiple of the uniform value. $p^\star$ is
the depth with the highest measured ratio and ``$p$ run'' the depth used in the main
runs.}
\label{tab:depth}
\begin{ruledtabular}
\begin{tabular}{cccccccc}
$n$ & $p=1$ & $p=2$ & $p=3$ & $p=4$ & $p=5$ & $p^\star$ & $p$ run \\
\hline
6 & 0.739 & 0.847 & 0.845 & 0.796 & 0.790 & 2 & 2 \\
& \textit{0.788} & \textit{0.930} & \textit{0.952} & \textit{0.943} & \textit{0.963} & & \\
& \small{7$\times$} & \small{12$\times$} & \small{12$\times$} & \small{11$\times$} & \small{11$\times$} & & \\[3pt]
8 & 0.659 & 0.781 & 0.784 & 0.772 & 0.747 & 3 & 3 \\
& \textit{0.692} & \textit{0.885} & \textit{0.932} & \textit{0.959} & \textit{0.968} & & \\
& \small{6$\times$} & \small{19$\times$} & \small{22$\times$} & \small{20$\times$} & \small{19$\times$} & & \\[3pt]
10 & 0.619 & 0.758 & 0.742 & 0.713 & 0.675 & 2 & 4 \\
& \textit{0.657} & \textit{0.902} & \textit{0.933} & \textit{0.946} & \textit{0.961} & & \\
& \small{8$\times$} & \small{27$\times$} & \small{28$\times$} & \small{25$\times$} & \small{24$\times$} & & \\[3pt]
12 & 0.588 & 0.727 & 0.710 & 0.671 & 0.645 & 2 & 5 \\
& \textit{0.609} & \textit{0.882} & \textit{0.918} & \textit{0.932} & \textit{0.952} & & \\
& \small{6$\times$} & \small{30$\times$} & \small{30$\times$} & \small{23$\times$} & \small{24$\times$} & & \\[3pt]
\end{tabular}
\end{ruledtabular}
\end{table}

Figure~\ref{fig:depth} and Table~\ref{tab:depth} show the expected shape. The noiseless
ratio rises with $p$ at every size and exceeds $0.95$ by $p = 5$. (At $n = 6$ it dips
very slightly between $p = 3$ and $4$, which is the multistart classical optimizer missing
the global optimum of an eight-parameter landscape rather than anything physical.) The
measured ratio rises steeply from $p = 1$, peaks, and then declines. The peak lies at
$p = 2$ or $3$ at every size; these are single $8192$-shot jobs, and at $n = 6$ and $8$
the top two depths differ by less than shot noise, so the data support the statement that
the optimum is at two or three layers but not which of the two. At $n = 12$ the measured
ratio falls by about $0.08$ between $p = 2$ and $p = 5$ while the noiseless value
continues to rise.

This partly separates the two effects mixed in Sec.~\ref{sec:results}. Comparing the
last two columns of Table~\ref{tab:depth}, the $n = 6$ and $8$ runs were at the device
optimum while the $n = 10$ and $12$ runs were past it. Part of the decline across
Fig.~\ref{fig:final} is thus a depth effect rather than a property of the instance, though
not all of it: even at each size's own best depth the measured ratio still falls with $n$.

The second measurement at $n = 12$ addresses a follow-up question. If the loss
comes from running too deep, could an optimizer that knew about the noise recover it by
choosing different parameters at the same depth? We reoptimized against the model of
Eq.~\eqref{eq:twin}, at the $\eta$ predicted by Eq.~\eqref{eq:etalaw}, and measured the
result. The noise-aware parameters differ from the noiseless-optimal ones by at most about
$0.01$ in measured ratio, in either direction, at every depth (orange curve,
Fig.~\ref{fig:depth}). At a given depth, in other words, reoptimizing against the noise model changed the
measured ratio by about $0.01$ at most. This is consistent with the picture in which the dominant error is a global depolarizing channel,
which contracts the whole output distribution toward uniform and therefore cannot be
steered around by a change of parameters; that noisy devices are limited in this way, and
that the limitation tightens with circuit volume, has been established in general
terms~\cite{stilckfranca2021limitations,wang2021noisebp}, of which this is a small
concrete instance.

One heuristic consistent with these data is to choose $p$ by maximizing the product of
the noiseless ratio and $1-\eta\big(3p\binom{n}{2}\big)$, with $\eta$ from
Eq.~\eqref{eq:etalaw}. For \ibmfez\ at these sizes it selects $p = 2$ at every size,
in line with the measured optimum lying at two or three layers. It requires an estimate of $\eta$ for the device in question.

\subsection{Repeatability across initializations}
\label{sec:replicates}

The comparison above is one trajectory per arm per size. To attach a spread to it we
repeated the $8$-asset experiment from three further starting points, changing nothing
else: same instance, same $p = 3$, same shots, same ten jobs per arm. The starts are the
linear ramp at $\Delta t \in \{0.35, 0.50, 0.60, 0.85\}$, with $\Delta t = 0.50$ the run
already reported; within a start both arms receive the same initial point.

\begin{figure*}[t]
\includegraphics[width=\textwidth]{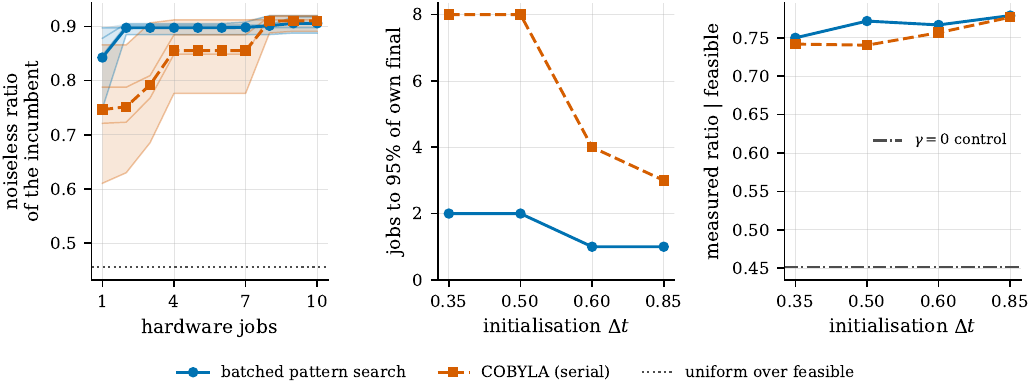}
\caption{Four independent initializations at $n=8$, $p=3$. \textbf{Left:} noiseless
approximation ratio of the incumbent against jobs; thin lines are individual starts,
markers the mean, bands the full range. \textbf{Centre:} jobs each arm needed to reach
$95\%$ of its own final value. \textbf{Right:} measured approximation ratio of the
read-out at each arm's returned parameters, against the $\gamma=0$ noise floor.}
\label{fig:replicates}
\end{figure*}

\begin{table*}[t]
\caption{The $8$-asset comparison from four initializations. ``COBYLA matches at'' is the
job at which the serial arm first reaches the noiseless quality the batched arm had after
its first job; ``best so far'' is the running maximum of the noiseless ratio at the end of
the budget and ``read-out'' the measured approximation ratio at the returned parameters.
$^{\dagger}$ marks the run reported above.}
\label{tab:replicates}
\begin{ruledtabular}
\begin{tabular}{cccccccccc}
& \multicolumn{2}{c}{after one job} & COBYLA & \multicolumn{2}{c}{jobs to 95\%} & \multicolumn{2}{c}{best so far} & \multicolumn{2}{c}{read-out} \\
$\Delta t$ & batched & COB & matches at & batched & COB & batched & COB & batched & COB \\
\hline
0.35\phantom{$^\dagger$} & 0.748 & 0.611 & 4 & 2 & 8 & 0.888 & 0.891 & 0.750 & 0.742 \\
0.50$^{\dagger}$ & 0.847 & 0.721 & 4 & 2 & 8 & 0.918 & 0.920 & 0.772 & 0.740 \\
0.60\phantom{$^\dagger$} & 0.878 & 0.788 & 4 & 1 & 4 & 0.920 & 0.919 & 0.767 & 0.757 \\
0.85\phantom{$^\dagger$} & 0.897 & 0.866 & 3 & 1 & 3 & 0.898 & 0.912 & 0.779 & 0.777 \\
\hline
mean & 0.843 & 0.746 & & 1.50 & 5.75 & 0.906 & 0.910 & 0.767 & 0.754 \\
\end{tabular}
\end{ruledtabular}
\end{table*}

The ordering holds at every start (Fig.~\ref{fig:replicates}, Table~\ref{tab:replicates}):
the batched search reaches $95\%$ of its own final value in one or two jobs at each of the
four starts, and COBYLA takes several more at each.

Two further features are worth noting. The final noiseless values of the two arms are
statistically indistinguishable, with means that differ by less than the standard
deviation of about $0.013$ across starts, which is the same message as before: the two methods find equally good
parameters and differ in how many round trips they take to do so. The device read-outs at
those parameters (Fig.~\ref{fig:replicates}, right) show a small edge for the batched arm
at each of the four starts, of the order of the spread, with both arms far above the
$\gamma=0$ floor. And on these four starts the advantage was largest at the poorest one, where COBYLA
takes longest to converge, and smallest at the best, where the ramp already lands near a
good region and both arms converge within a few jobs. If that holds more generally, a
practitioner who cannot pre-tune the initialization --- which is the situation whenever the
instance is too large to simulate --- is the one for whom the batched method would help
most.

\section{Discussion}
\label{sec:discussion}

\subsection{What the comparison does and does not show}

The batched search does not find better optima than COBYLA in any reliable sense. Measured
by the best point each arm reached, the two end within about $0.01$ of each other at every
size; measured by the parameters each arm actually returned, the two are level at $n = 6$
and $8$ and the batched arm is ahead by a few hundredths at $n = 10$ and $12$
(Table~\ref{tab:convergence}). What the batched search does is reach comparable
parameters in fewer jobs, and on the access mode of Sec.~\ref{sec:cost} jobs were the
dominant cost. The claim is about the shape of the convergence curve near its origin,
not about its asymptote, and the regime it speaks to is the one in which a user has time
for a handful of round trips rather than a hundred.

Several limitations bound the strength of the evidence, and we list them here.

\emph{Replication.} The four main instances were each run once. The replicate study of
Sec.~\ref{sec:replicates} attaches a spread to the $8$-asset comparison, but the $10$- and
$12$-asset results are single trajectories, and the differences in final quality there
should be read against the replicate standard deviation of about $0.013$ rather than as
precise.

\emph{Sequential arms.} The two arms ran one after the other rather than interleaved. The
drift control of Sec.~\ref{sec:protocol} is a single paired repeat per instance, which
cannot separate drift from shot noise; it bounds the effect at a few hundredths of a
$\CVaR$ unit and no more than that. Interleaving the arms would have been cleaner and
would have cost nothing.

\emph{Ablations not run.} Two further arms would sharpen the attribution and each would
cost one job budget. A random-batch arm, with $M$ points per job drawn around the same
start and tracked greedily, would separate the value of batching itself from the value of
the pattern-search stencil. A serial pattern search, one point per job with the same
acceptance rule, would separate the stencil from the shot-noise tolerance of Algorithm~1,
which is a noise-robustness feature COBYLA has no counterpart of. As designed, the
experiment measures the package rather than its parts.

\emph{A batchable competitor.} COBYLA was chosen because it was the strongest serial
method in the simulations, but as noted in Sec.~\ref{sec:intro}, SPSA and parameter-shift
gradients are themselves batchable, as are model-based trust-region
methods~\cite{sung2020models}, multistart schemes~\cite{shaydulin2019multistart} and
information-sharing schemes~\cite{self2021infosharing}. A batched SPSA, or several
concurrent COBYLA chains packed into one PUB, would be the harder comparison, and we did
not run it.

\emph{The oracle metric.} The per-job curves of Fig.~\ref{fig:convergence} use the exact
noiseless ratio, which no optimizer can compute and which does not exist at a problem size
worth solving. It diagnoses the search; it is not a quantity either method returns. The
device read-outs of Tables~\ref{tab:final} and \ref{tab:replicates} are the non-oracle
counterpart, and there the two arms are closer.

\emph{Selection bias in the measured traces.} The lower row of Fig.~\ref{fig:convergence}
plots a running minimum of measured $\CVaR$. The batched arm takes that minimum over $M$
values per job and the serial arm over one, so the batched trace carries a downward
selection bias. This is one reason the separation is larger there than in the
independently measured read-outs, and the read-outs are the numbers to rely on.

\emph{Problem size.} Twelve qubits is far below the size at which any quantum method could
matter for portfolio selection, and the instances here are solved by enumeration in
milliseconds. The sizes were chosen so that exact answers were available for every
check in the paper. Nothing here bears on whether QAOA can compete with classical solvers
at larger sizes.

\subsection{Depth and instance size were varied together}

The circuit depth was increased with the instance size, so the decline in measured
quality across Fig.~\ref{fig:final} mixes a depth effect with a size effect, and the depth
scan of Sec.~\ref{sec:depth} only partly separates them. Repeating the $n = 12$
optimization at $p = 2$ rather than $p = 5$ is the obvious next experiment;
Table~\ref{tab:depth} suggests it would raise the final read-out, and it would not change
the job-efficiency comparison, which is a statement about the optimizer at a fixed
circuit.

\subsection{Relevance beyond this experiment}

The cost structure of Sec.~\ref{sec:cost} is a property of the access mode rather than of
QAOA. Any variational procedure driven through a queue --- an eigensolver for
chemistry~\cite{kandala2017vqe}, a quantum kernel evaluation, a calibration or
error-mitigation sweep --- pays the same fixed charge per round trip and can, in
principle, fill the same PUB. Three suggestions follow.

First, when comparing optimizers for use on cloud hardware, report convergence against
submitted jobs as well as against function evaluations. The two plots can look quite
different, as Fig.~\ref{fig:convergence} shows, and the job axis is the one a user can
translate into time and cost. Application-oriented benchmarking
efforts~\cite{lubinski2023benchmarks} that already report execution time including cloud
overheads are a natural place for such an axis.

Second, among optimizers of comparable sample efficiency, prefer those whose next batch
of query points is known in advance. The property admits degrees --- pattern search names
$4p+2$ points per round trip, parameter-shift gradients $4p$, SPSA and its quantum-natural
variant~\cite{gacon2021qfi} two, and COBYLA, Nelder--Mead and Powell one --- and an
optimizer that is slightly worse per evaluation but names many points at a time can be
considerably better per job.

Third, state the cost model alongside the result. Shots are reported because they were
once the binding constraint; on the access mode used here they were not, and a paper that
reports only shots leaves the reader unable to reconstruct what the experiment would cost
them.

We should also be clear about what a job-budget framing does not do. Nothing here
addresses the fundamental obstacles to variational scaling, barren
plateaus~\cite{mcclean2018barren} and their noise-induced
counterpart~\cite{wang2021noisebp}, and the narrowing separation between the measured
curves and the noise floor in Fig.~\ref{fig:final} is a reminder that at twelve qubits and
nearly a thousand two-qubit gates most of the output distribution is already uniform. Job
efficiency affects how quickly the device's output quality is reached, not that quality.

\section{Conclusion}

We compared a batched pattern search with COBYLA for optimizing QAOA parameters on IBM's
\ibmfez\ processor, with the two methods given the same number of submitted jobs and the
same number of shots, on four nested portfolio-selection instances small enough to be
solved exactly. The batched search, which evaluates its whole iteration in one job,
reached a good parameter region after a single job at every size, where COBYLA needed
several; the two methods converged to comparable final quality, and the ordering was
unchanged across four starting points at the $8$-asset size. The device's output at the
returned parameters was correlated with the exact ranking of portfolios and clearly
separated from a control circuit of the same depth with the cost layer removed. A depth
scan at classically optimal parameters located the device's optimal circuit depth at two
or three layers, beyond which additional layers reduced the measured quality, and
reoptimizing the parameters against a noise model did not recover the loss.

The optimizer is deliberately simple. The observation we would draw from the exercise
is that, when quantum processors are reached through a queue, the number of times an
algorithm goes around the loop is a cost worth counting, and that counting it can change
the choice of classical optimizer. We have tried to describe
the experiment in enough detail, and to release enough of its material, that others can
repeat it or improve on it.

\begin{acknowledgments}
The author thanks IBM Quantum for access to \ibmfez. This work was carried out within the
$180$ minutes of quantum processing time granted under the IBM Quantum Open Plan promotion
of March 2026, of which about $34$ minutes (by the estimate of Sec.~\ref{sec:cost}) were
used. The author is a member of the Qiskit Advocate Program. The views expressed are those
of the author and do not reflect the official policy or position of IBM or the IBM Quantum
team.
\end{acknowledgments}

\section*{Data and code availability}

Every hardware result in this paper is produced by six self-contained Jupyter notebooks,
one per instance size plus the replicate and depth-scan studies, which are released
together with the raw data they produced: the complete job log including every IBM job
identifier, the measured bitstring counts of every parameter row of every job, both
optimization traces, the control read-outs and the final metrics. The figures and tables
are generated from those files by the included scripts. The archive is available at
\url{https://github.com/muf148/qaoa-job-budget}; code is released under the MIT licence and
data under CC-BY-4.0.

\appendix

\section{Choosing the hyperparameters in simulation}
\label{app:tuning}

Every hyperparameter was fixed on a noise model of \ibmfez\ before any hardware time was
used. Full noisy simulation is too slow to sweep --- one $12$-qubit, $p = 2$ circuit at
$2048$ shots takes over a minute with a full Aer noise model --- so the sweep used the
one-parameter surrogate of Eq.~\eqref{eq:twin}, with $\eta$ fitted for each $(n,p)$
against Aer at two parameter points. The surrogate reproduced Aer's $\CVaR$ to about $0.1$
in normalized units and its feasible mass to about $0.05$.

The sweep covered $p \in \{1,2,3\}$, shots per candidate $\in \{512, 1024, 2048\}$, jobs
per arm $\in \{8,12,16\}$, $\alpha \in \{0.1, 0.25, 0.5\}$ and initial radius
$\in \{0.25, 0.4\}$, restricted to configurations fitting a $300$\,s estimated budget, at
ten random seeds each, followed by a confirmation study of the selected settings at forty
seeds. It selected $\alpha = 0.5$ at every size and an initial radius of $0.25$, and found
the final quality insensitive to shots per candidate over the range swept and insensitive
to the number of jobs beyond about ten. The shots and jobs actually used
(Table~\ref{tab:ledger}) were then set by hand within this guidance and within the
hardware time available; the $300$\,s cap applied to the sweep, not to the final runs. On
the surrogate the batched search was ahead of COBYLA at every job budget swept, with COBYLA
needing sixteen jobs to match what the batched search had after twelve at $n = 6$, $8$ and
$10$, and not matching it within sixty-four jobs at $n = 12$. The hardware runs show the
same ordering under the different metric of Table~\ref{tab:convergence}, though the
surrogate carries neither crosstalk nor drift and its absolute numbers should not be
compared with the device's.

The surrogate selected $p = 2$ at all four sizes, on the grounds that $p = 3$ improved the
predicted ratio only slightly for $50\%$ more two-qubit gates. The depths actually run
were $p = 2,3,4,5$, chosen by hand in the belief that a larger instance warrants a deeper
circuit. The hardware depth scan of Sec.~\ref{sec:depth} was consistent with the surrogate's
recommendation: the measured optimum is at two or three layers at every size. A fake backend
carries no crosstalk and no drift, which is why the $\gamma = 0$ and drift controls of
Sec.~\ref{sec:protocol} are part of every hardware run; but on the one question where its
recommendation was overruled, it should have been followed.

\section{Reproducing the experiment}
\label{app:repro}

Each notebook is self-contained and runs top to bottom against \ibmfez\ in roughly $100$
to $600$\,s of estimated metered time. A \texttt{DRY\_RUN} switch replaces the device with
the surrogate of Eq.~\eqref{eq:twin} and executes the entire experiment in about a minute
at no cost, which is the recommended first step. A ledger is enforced at run time: before
every submission the notebook predicts the metered cost from Eq.~\eqref{eq:cost} and
refuses to submit a job that would exceed the configured cap.

Four implementation details caused errors that were silent rather than loud, and we record
them because they would cost another user a run.
\begin{enumerate}
\item \texttt{BitArray.get\_int\_counts()} called with no argument merges the counts of
      all parameter rows of a PUB into one dictionary. A batched run that does this
      produces plausible-looking output that is meaningless. The row index is mandatory.
\item The parameters of a transpiled circuit are ordered by name, not by construction
      order. The array of parameter values must be assembled by name lookup.
\item An explicit \texttt{initial\_layout} must be checked to have survived the pass
      manager; otherwise the circuit can silently move off the chosen chain.
\item Reading the runtime's reported usage through \texttt{job.metrics()} is
      version-dependent, and a \texttt{try/except} around it will hide a wrong key path
      behind a plausible fallback. Ours did, on every job of this study, which is why the
      metered times reported in Sec.~\ref{sec:cost} are estimates rather than
      measurements. If the device's own number is wanted, the code should check that it
      actually received one.
\end{enumerate}
The first and the last of these have the same shape, a fallback that produces output
indistinguishable from success, and we would encourage anyone building this kind of
pipeline to make both fail loudly.

\bibliographystyle{apsrev4-2}
\bibliography{refs}

\end{document}